\documentclass[12pt]{article}

\usepackage{amsmath,epsfig,epsf,psfrag,natbib}
\usepackage{amssymb}
\usepackage{amsbsy}
\usepackage{lscape}
\usepackage{color,latexsym,amsfonts,amsthm,amscd,graphicx}
\usepackage{mathrsfs}
\usepackage{amsfonts}
\usepackage{amsmath}
\usepackage{color}
\usepackage{natbib, url}
\usepackage{mathrsfs}  %   $\mathscr{F}$
\usepackage{amsmath,amssymb}
\usepackage{subfigure}
\usepackage{graphicx}
\usepackage{epstopdf}
\usepackage{hyperref}
\usepackage{bm,bbm}
\usepackage{booktabs}
\usepackage{multirow}
\usepackage{mathrsfs}
\usepackage{float}
\usepackage{ragged2e}
\usepackage{adjustbox}
\usepackage[figuresright]{rotating}
\usepackage{appendix, natbib}
\usepackage{algorithmicx,algorithm,algpseudocode}
\usepackage{comment}
\usepackage{mathrsfs}
\usepackage{dsfont}
\algnewcommand{\Input}{\item[\textbf{Input:}]}
\algnewcommand{\Output}{\item[\textbf{Output:}]}
\algnewcommand{\Initial}{\item[\textbf{Initialization:}]}

\newcommand{\be}{\begin{equation}}
\newcommand{\ee}{\end{equation}}
\newtheorem{theorem}{Theorem}
\newtheorem{lemma}{Lemma}

\newtheorem{proposition}{Proposition}

\newtheorem{assumption}{{\sc Assumption}}

\newcommand{\bpi}{ {\bm \pi}}
\newcommand{\bzero}{ {\bf 0}}
\newcommand{\diag}{ {\rm diag}}

\newcommand{\bI}{ {\bf I}}

\newcommand{\bba}{ {\bm a}}

\def\bT{{\boldsymbol\Theta}}
\def\bt{{\boldsymbol\theta}}
\def\bbeta{{\boldsymbol\beta}}
\def\bphi{{\boldsymbol\phi}}
\def\bga{{\boldsymbol\gamma}}
\def\bSigma{{\boldsymbol\Sigma}}

\newcommand{\bQ}{ {\bf Q }}
\newcommand{\bW}{ {\bf W }}
\newcommand{\bS}{ {\bf S}}
\newcommand{\blambda}{ {\bm \lambda}}

\def\bT{{\boldsymbol\Theta}}

\newcommand{\ba}{\begin{eqnarray}}
\newcommand{\ea}{\end{eqnarray}}
\newcommand{\bas}{\begin{eqnarray*}}
\newcommand{\eas}{\end{eqnarray*}}
\newcommand{\bit}{\begin{itemize}}
\newcommand{\eit}{\end{itemize}}
\newcommand{\ben}{\begin{enumerate}}
\newcommand{\een}{\end{enumerate}}
\makeatletter
\newcommand{\Rmnum}[1]{\expandafter\@slowromancap\romannumeral #1@}
\makeatother

\newcommand{\e}{ { \mathbb{E}}}

\def\T{{ \mathrm{\scriptscriptstyle \top} }}

\newcommand{\bx}{ \boldsymbol{x}}

\newcommand{\bX}{ \boldsymbol{X}}
\usepackage{authblk}
\begin{document}
\date{}

\title{Semiparametric Learning from Open-Set Label Shift Data}

\author[1]{Siyan Liu}
\author[2]{Yukun Liu\thanks{Corresponding author: ykliu@sfs.ecnu.edu.cn}}
\author[3]{Qinglong Tian}
\author[3]{Pengfei Li}
\author[4]{Jing Qin}
\affil[1,2]{
    KLATASDS-MOE,
 School of Statistics,
  East China Normal University,
    Shanghai 200062, China}
\affil[3]{Department of Statistics and Actuarial Science, University of Waterloo,
Ontario N2L 3G1, Canada}
\affil[4]{National Institute of Allergy and Infectious Diseases,
 National Institutes of Health, Rockville, MD 20892, U.S.A.}
\renewcommand*{\Affilfont}{\small }
\renewcommand\Authands{ and }
\date{}
\maketitle

\abstract{
We study the open-set label shift problem, where the test data may include a novel class absent from training. This setting is challenging because both the class proportions and the distribution of the novel class are not identifiable without extra assumptions. Existing approaches often rely on restrictive separability conditions, prior knowledge, or computationally infeasible procedures, and some may lack theoretical guarantees.
We propose a semiparametric density ratio model framework that ensures identifiability while allowing overlap between novel and known classes. Within this framework, we develop maximum empirical likelihood estimators and confidence intervals for class proportions, establish their asymptotic validity, and design a stable Expectation–Maximization algorithm for computation. We further construct an approximately optimal classifier based on posterior probabilities with theoretical guarantees.
Simulations and a real data application confirm that our methods improve both estimation accuracy and classification performance compared with existing approaches.
}


% \boxedtext{
% \begin{itemize}
% \item Key boxed text here.
% \item Key boxed text here.
% \item Key boxed text here.
% \end{itemize}}

\maketitle

\allowdisplaybreaks
\section{Introduction}
\subsection{Open-Set Label Shift Problem}

\begin{figure}[!ht]
  \centering
  \includegraphics[width=0.85\textwidth]{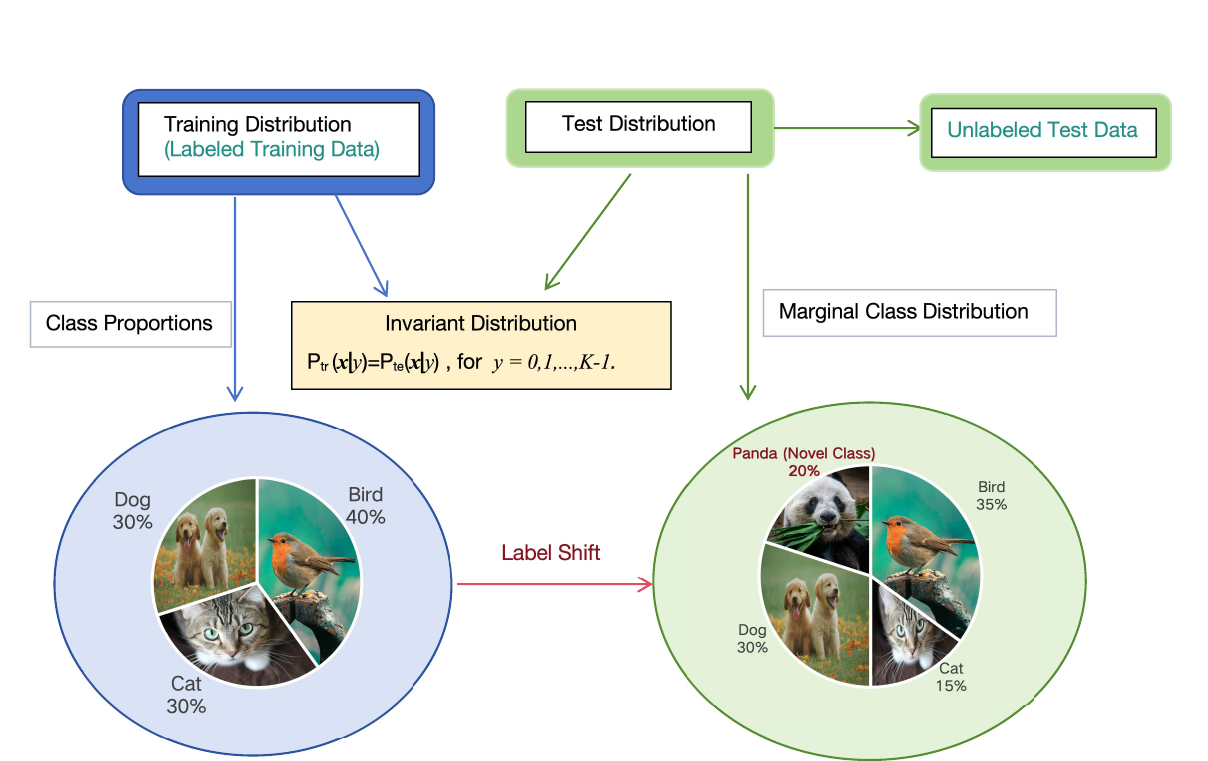}
   % \vspace{-6pt}
  \caption{
  Schematic overview of OSLS setting.
       }
  \label{Fig-OSLS}
  \end{figure}

Consider a multi-class classification task with response variable 
\(Y \in \{0,1,\dots,K\}\) and covariates \(\bX\). 
In the open-set label shift (OSLS) problem, the class \(Y = K\) 
is defined as the novel class because it appears only in the test data 
but not in the training data.
Specifically, we observe a labeled training data
\(
\mathcal{L} = \{( \boldsymbol{x}_i, y_i)\}_{i=1}^n,
\)
where $y_i\neq K$,
and an unlabeled  test data 
\(
\mathcal{U} = \{\boldsymbol{x}_{n+j}\}_{j=1}^m,
\)
where some test labels may equal
$K$. Throughout this paper,   we assume  $n_k = \sum_{i=1}^n I(y_i =k)$
is positive and
adopt a retrospective sampling scheme for the training data:
for each class $k=0,1,\dots , K-1$, 
the sample size $n_k$ is fixed in advance, and
the covariates of the $n_k$ instances with label $k$ in the training data are drawn  from the conditional distribution of
 $\bX$  given  $Y = k$.  
For the known classes, we assume distributional invariance between the training and test data  \citep{garg2022domain}:
the conditional distribution of $\bX$ in the training data,
denoted $P_{\text{tr}}(\bx | y)$,
is identical to that in the test data,
denoted $P_{\text{te}}(\bx | y)$, i.e.,
\begin{equation}
\label{invariance}
 P_{\text{tr}}(\bx | y) = P_{\text{te}}(\bx | y), \quad y = 0, 1, \ldots, K-1.
\end{equation}
Let  $\pi_k$ denote the proportion of test observations belonging to class $k$, $k=0,1,\ldots,K$.
We allow for label shift among the known classes; that is, the ratio
$
{n_j}/{n_k}$    may differ   from ${ \pi_j}
/{\pi_k}$ for some \(j\ne k\) in \(\{0, 1, \dots, K-1\}\).
A schematic overview of the OSLS setup is shown in Figure~\ref{Fig-OSLS}. Our objective is to make inference on
$\pi_k$ for $k=0,1,\ldots,K$ and to classify the test observations under this setting.

The OSLS framework has gained growing attention in recent years due to its relevance in many real-world applications \citep{garg2022domain}. For example, in a facial recognition system trained on labeled data of authorized personnel for secure access control \citep{1512050}, the deployed system inevitably encounters unlabeled inputs that include not only known individuals but also visitors or intruders absent from training. Another important example is species distribution modeling in ecology using presence-only data \citep{ward2009presence}, where a sample of confirmed presence records (e.g., from field surveys) is available, together with an unlabeled sample from the broader study region that contains both presence and absence instances. This setting corresponds to positive-unlabeled (PU) learning \citep{elkan2008learning,blanchard2010semi,scott2015rate,liu2025positive}, a special case of OSLS with
$K=1$.
\subsection{Challenges and Related Work\label{challenges_related_work}}
Under the OSLS framework, making inference on
$\pi_k$
and classifying test observations is challenging, and in fact statistically infeasible, because the distribution of $\bX$ in the novel class $K$ and $\pi_k$'s are not identifiable without additional assumptions.
To see this, define $f_k(\bx)=P_{\text{te}}(\bx|Y=k)$ for $k=0,\ldots,K$.
Then
\begin{equation}
\label{test.model}
\{\bx_{n+j}\}_{j=1}^m
\sim P_{\text{te}}(\boldsymbol{x}):=\sum_{k=0}^K \pi_k f_k(\bx).
\end{equation}
Under \eqref{invariance},
$\{f_k(\bx)\}_{k=0}^{K-1}$  can be identified from the labeled training data.
However, $\pi_k$'s and $f_K(\bx)$ are not identifiable, because for any $\rho\in(0,1)$,
\begin{equation}
\label{eq_identifiability}
\sum_{k=0}^K \pi_k f_k(\bx)
=\sum_{k=0}^{K-1}
\pi_k^* f_k(\bx) +\pi_K^*\cdot f^*_K(\bx),
\end{equation}
where $\pi_k^*=\rho\pi_k$ for $k=0,\ldots,K-1$, $\pi_K^*=\{1-\rho(1-\pi_K)\}$, and
$$
f^*_K(\bx)=\frac{
\sum_{k=0}^{K-1}
(1-\rho)\pi_k f_k(\bx)+\pi_K f_K(\bx)}
{1-\rho(1-\pi_K)}.
$$
In other words, $(\pi_0,\ldots,\pi_K, f_0,\ldots,f_{K-1},f_K)$ and
$(\pi_0^*,\ldots,\pi_K^*, f_0,\ldots,f_{K-1},f_K^*)$ both correspond to the same  $P_{\text{te}}(\boldsymbol{x})$.

In the literature, several classes of methods have been proposed to address the non-identifiability issue.
The first class assumes that the $\pi_k$'s are fully known. Methods in this category have been developed for both the binary case ($K = 1$, i.e., the PU learning problem) \citep{steinberg1992estimating, ward2009presence, song2020pulasso} and the multi-class setting ($K > 1$)
\citep{xu2017multi,zheng2023high}.
Although this assumption renders all $f_k$ identifiable, our numerical studies in Section \ref{Numerical_studies} demonstrate that misspecifying the $\pi_k$'s can severely degrade classification performance.

The second class of methods addresses non-identifiability by imposing separability conditions. These range from the strict no-overlap assumption between novel and existing classes \citep{elkan2008learning, du2014class, northcutt2017learning}, to the more relaxed anchor set assumption  \citep{scott2015rate, liu2015classification, bekker2018estimating}, and further to the positive subdomain assumption \citep{ramaswamy2016mixture, guan2022conformal}. Approaches in this category have been studied extensively for  $K=1$ (see \citeauthor{zhu2023mixture}, \citeyear{zhu2023mixture} for a review) and have more recently been extended to $K>1$ \citep{garg2022domain}.
Although \citet{garg2022domain} established the identifiability of model parameters under certain separability conditions and proposed the PULSE method for estimating $\pi_k$'s and classifying test observations simultaneously, their approach has two main limitations. First, the separability conditions are designed primarily for theoretical identifiability but are difficult to enforce in practice. They can also be restrictive, as they are not satisfied by many commonly used distributions such as the normal distribution. When these conditions are violated, the PULSE method may produce biased estimates of $\pi_k$'s and suffer a substantial loss in classification performance (see Section \ref{Numerical_studies} for details). Second, the PULSE method does not provide inference procedures, such as confidence intervals for $\pi_k$'s.

The third class of methods relies on an irreducibility condition, which requires that $f_k(\bx)$ cannot be expressed as a mixture of $\{{f_l(\bx)}\}_{l\neq k}$ and another distribution for $k=0,\ldots,K$. This condition is weaker than the separability assumptions in the second class of methods.
Most theoretical progress in this direction has focused on the case $K=1$ \citep{blanchard2010semi,jain2016estimating,ivanov2020dedpul}. In particular, \citet{blanchard2010semi} introduced the notion of irreducibility, established identifiability under this condition, and proposed a consistent estimator for $\pi_1$. However, their estimator is computationally infeasible \citep{garg2021mixture}.
Extending to $K>1$, \citet{sanderson2014class} reformulated the OSLS problem as $K$ separate PU learning problems, applying \citet{blanchard2010semi}'s method to estimate each $\pi_k$ for $k=0,\ldots,K-1$. This approach has three limitations: (i) it inherits the computational intractability of \citet{blanchard2010semi}'s estimator, (ii) estimation errors may accumulate across the $K$ PU problems, reducing efficiency \citep{garg2022domain}, and (iii) classification of test observations was not formally addressed.
 Several numerical methods from open-set domain adaptation also fall under this class, see \citet{NEURIPS2020_bb7946e7} and references therein.
%\citep{Busto2017ICCV,Saito_2018_ECCV,You2019CVPR,Tan_2019_CVPR,NEURIPS2020_bb7946e7,10.1007/978-3-030-58555-6_34}.
These approaches, however, are largely heuristic, lack theoretical guarantees \citep{garg2022domain}, and focus on classifying test observations rather than estimating the $\pi_k$'s.
In summary, within the third class, no existing method with theoretical support can simultaneously estimate the $\pi_k$'s and classify test observations when $K>1$.

\subsection{Our Contributions and Overview}

We address the non-identifiability issue in \eqref{test.model} using the semiparametric density ratio model (DRM; \citeauthor{anderson1979multi}, \citeyear{anderson1979multi}; \citeauthor{qin2017biased}, \citeyear{qin2017biased}).
This framework allows overlap between the novel and known classes, and eliminates the need for separability conditions or prior knowledge of the $\pi_k$'s in the test data. Our contributions are summarized as follows.

\begin{enumerate}
  \item We formally establish that all model parameters in \eqref{test.model}, including the $\pi_k$'s, are identifiable under the proposed semiparametric framework.

  \item We estimate the model parameters using the maximum empirical likelihood method and further demonstrate the asymptotic normality of the resulting estimators.

  \item We present a numerically stable Expectation-Maximization (EM) algorithm for implementation and verify its monotonicity.

  \item We construct empirical likelihood ratio based confidence intervals for each $\pi_k$ in the test data, for $k = 0, \ldots, K$. To our knowledge, these are the first statistically valid confidence intervals for $\pi_k$ under the OSLS setting.

  \item We design an approximately optimal classifier under a cost-sensitive loss function for the test data. The classifier relies on posterior probabilities, which have closed-form expressions in terms of the model parameters. Consequently, it achieves more reliable performance compared to existing methods in the OSLS setting.
\end{enumerate}

The remainder of this paper is organized as follows. In Section \ref{Model_Setup_and_Identifiability}, we introduce the model setup and establish the identifiability of all underlying parameters. Section \ref{drm-EL} presents the maximum empirical likelihood method, the EM algorithm for numerical implementation, and the asymptotic properties. Section \ref{Classification_under_SPEL_Framework} addresses the classification problem in the test data. Section \ref{Numerical_studies} evaluates the empirical performance of the proposed methods through simulation studies and a real data application. Finally, Section \ref{Discussion} concludes the paper with a discussion. For clarity, all proofs are provided in the supplemental materials.

\section{Identifiability under Density Ratio Model}
\label{Model_Setup_and_Identifiability}

In this section, we address identifiability in \eqref{test.model}.
  Recall from \eqref{eq_identifiability} that $f_K$ and $\pi_k$'s cannot be identified without additional assumptions on $f_K$, even when $\{f_k\}_{k=0}^{K-1}$ are known. To avoid the inflexibility of fully parametric models for $f_K$ while still leveraging shared structure across classes, we assume that the $f_k(\bx)$'s follow a semiparametric DRM:
  \ba
  \label{drm}
  f_k(\bx) = f_0(\bx) e^{ \alpha_k+\bbeta_k^\T \bphi(\bx) }, \quad k= 1, \; 2 \dots , K,
  \ea
  where $\bphi(\bx)$ is a pre-specified
  $q$-dimensional vector-valued function of $\bx$, $(\alpha_k,\bbeta_k)$ are unknown model parameters, and the baseline density $f_0(\bx)$ is unspecified, making the DRM semiparametric.
A common choice for $\bphi(\bx)$ is simply $\bx$. Polynomial functions of $\bx$ can also be used to increase model flexibility. For image data, $\bphi(\bx)$ may be taken as the embedding layer of a pre-trained neural network.

As a semiparametric model, the DRM combines the interpretability of parametric models with the flexibility of nonparametric methods. It is commonly used in closed-set distribution shift problems to model the probabilistic relationships between training and test data \citep{Shimodaira2000improving, Sugiyama2007,pmlr-v80-lipton18a}.

Let  $\bga_k = (\alpha_k, \bbeta_k^\T)^\T$ 
for $k=1,\ldots, K$ and $\bphi_e(\bx) = (1, \bphi^\T(\bx))^\T$.
  We set $\bga_0=\bzero$ for notational simplicity.
Under model \eqref{drm}, $P_{\text{te}}(\boldsymbol{x})$ in \eqref{test.model}
can be written as
\ba
\label{test.model.drm}
P_{\text{te}}(\boldsymbol{x})
=
f_0(\bx) \left\{
 \sum_{k=0}^K \pi_k e^{\bga_k^\T \bphi_e(\bx) } \right\}
 =
f_0(\bx) \left[
1+
 \sum_{k=1}^K \pi_k \left\{e^{\bga_k^\T \bphi_e(\bx) }
 -1\right\}\right].
 \ea
We show that under mild conditions,
the underlying parameters in
$P_{\rm te}(\bx) $ are identifiable
based on the training data and test data.
Throughout this paper, we use a superscript ``o'' to highlight the true value of a generic parameter,
e.g., $\bbeta_1^{o}$ denotes the true value of $\bbeta_1$, and we
 use $\e_0$ to denote
 the expectation operator with respect to the baseline density $f_0(\bx)$.
 \begin{assumption}
  \label{ass:ass1}
  Let $N=n+m$, and
   $n_k / N = c_k$ for $k = 0, 1, \dots, K-1$, where each $c_k \in (0, 1)$ is a constant. Furthermore, $c = m/N$ is also a constant in $(0, 1)$.
\end{assumption}

\begin{assumption}
  \label{ass:ass2}
(i)  $\bbeta_k^o \ne \bzero ,\ \bbeta_i^o \ne \bbeta_j^o$, for $i \ne j$, $1 \le i, j, k \le K$.
(ii) $\pi_K^o >0$.
(iii)   $\e_0\{  \bphi_e(\bX) \bphi_e^\T  (\bX) \}$ is finite and positive definite.

\end{assumption}

Assumption \ref{ass:ass1} ensures that the sample sizes for the $K$ known classes in the training data, as well as the overall training and test sample sizes, are of the same order. This assumption can be relaxed to allow
$n_k / N \to c_k$ for $k=0,1,\ldots,K-1$ and $m/N \to c$ as $N \to \infty$, but for simplicity and clarity, we take $c_k$'s and $c$ as fixed constants under Assumption \ref{ass:ass1}. This simplification does not affect the technical conclusions.
Assumption \ref{ass:ass2} typically holds when all $K+1$ densities are distinct and the proportion of the novel component is non-negligible. Moreover, the condition that $\e_0\{\bphi_e(\bX) \bphi_e^\T(\bX)\}$ is nonsingular in Assumption \ref{ass:ass2} ensures the identifiability of $\bbeta_k$.

 Denote
 $\bpi=(\pi_1,\pi_2, \dots , \pi_K)^{\T}$,
 $\bga=(\bga_1^{\T},\bga_2^{\T} , \dots , \bga_K^{\T})^{\T}$,
 and
 $\bt=(\bga^\T,\bpi^\T)^\T$.
 The following lemma establishes the identifiability of the model parameters in \eqref{drm}.
 %Its proof is provided in Section 2 of the supplementary material.

\begin{lemma}
  \label{identifiability}
  Under Assumptions   \ref{ass:ass1} and \ref{ass:ass2},  $f_0(\bx)$ and $\bt $  are   identifiable.
\end{lemma}

Under \eqref{drm}, Lemma \ref{identifiability} implies that all $\pi_k$ and $f_k$ are identifiable; consequently, all model parameters in \eqref{test.model} are also identifiable.

\section{Maximum Empirical Likelihood Method \label{drm-EL}}

\subsection{Empirical Likelihood}

For convenience, let  $F_k$ denote  the cumulative distribution function (cdf)  of $f_k$ for $k=0,\ldots,K$.
Under model \eqref{drm}, the likelihood contribution of the training data  is
\begin{align}
\label{likelihood.L0}
L_0 &=
\prod_{i=1}^n \prod_{k=0}^{K-1} \{ d F_k(\bx_i)\}^{I(y_i=k)}
=
\prod_{i=1}^n \prod_{k=0}^{K-1} \{  e^{ \bga_k^\T \bphi_e(\bx_i)  }  dF_0(\bx_i)\}^{ I(y_i=k) }.
\end{align}
Using \eqref{test.model.drm},
  the likelihood contribution of the testing data $\{\bx_{n+j}\}_{j=1}^m$ is
\begin{align}
\label{likelihood.L1}
L_1 &=  \prod_{i=n+1}^{N}  \left[   1+ \sum_{k=1}^K
 \pi_k  \{ e^{\bga_k^\T \bphi_e(\bx_i) } -1 \} \right]  dF_0(\bx_i).
\end{align}
Define $D_{i} = 0$ for $ 1\leq i\leq n$ and $D_i=1$ for $n+1\leq i\leq N$.
Combining \eqref{likelihood.L0}-\eqref{likelihood.L1},
we have the full likelihood $L_0  L_1$ or equivalently 
\begin{align}
     \prod_{i=1}^{N} \left\{ dF_0(\bx_i)   \prod_{k=0}^{K-1}   e^{  (1-D_i ) \bga_1^\T \bphi_e(\bx_i)  I(y_i=k) }
     \left[
  1+ \sum_{k=1}^K
  \pi_k  \{ e^{\bga_k^\T \bphi_e(\bx_i) } -1 \}  \right]^{D_i} \right\}.\label{full.likelihood}
\end{align}

We use empirical likelihood (EL;  \citeauthor{Owen2001}, \citeyear{Owen2001}) to handle the nonparametric baseline distribution $F_0$. Following the EL principle, $F_0$ is modeled as a discrete distribution
$
F_0(\bx) = \sum_{i=1}^{N} p_i I(\bX_i\leq \bx),
$
where $p_i=dF(\bx_i)$, $i=1,\ldots,N$.
Substituting $p_i = dF(\bx_i)$ into \eqref{full.likelihood} and taking the logarithm, we have the log-EL
\begin{align}
\nonumber
\tilde \ell =&
\sum_{i=1}^{N} \left\{  \sum_{k=1}^{K-1}  \bigl(1-D_i\bigr) \bga_k^\T \bphi_e(\bx_i) I(y_i=k)
+    D_i \log\left[ 1 +  \sum_{k=1}^K \pi_k \{e^{\bga_k^\T \bphi_e(\bx_i)}-1 \} \right]\right\}\\
&+\sum_{i=1}^N \log (p_i),\label{log_likelihood_func}
\end{align}
where feasible $p_i$'s satisfy
\ba
\label{constraints}
p_i\geq 0, \quad\sum_{i=1}^N p_i = 1, \quad
\sum_{i=1}^N p_i \{ e^{\bga_k^\T \bphi_e(\bx_i)} - 1 \} = 0, \quad k=1, 2, \dots , K .
\ea
The first two constrains in \eqref{constraints} ensures that $F_0$ is a valid cdf, while the last set of constraints ensures that $F_k$ for $k=1,\ldots,K$ are also valid cdfs.

Inferences about the underlying parameters are typically made through their profile log-EL function. Given $\bt$, the log-EL $\tilde \ell$ is maximized with respect to $p_i$ under the constraints in \eqref{constraints} at
\ba
\label{p-mle}
  p_i = \frac{1}{N} \frac{1}{1+ \sum_{k=1}^K\lambda_k \{ e^{\bga_k^\T \bphi_e(\bx_i)} - 1 \}},
\ea
where $ \{\lambda_k\}_{k=1}^K $ solves
\ba
\label{equ-lambda12}
\frac{1}{N} \sum_{i=1}^{N} \frac{  e^{\bga_k^\T \bphi_e(\bx_i)} - 1}{1+ \sum_{k=1}^K\lambda_k \{ e^{\bga_k^\T \bphi_e(\bx_i)} - 1 \}} =  0 , \ k= 1,2,\dots, K.
\ea
Accordingly, up to a constant independent of $\bt$, the profile log-EL function of $\bt$ (after maximizing over $p_1, \ldots, p_N$) is
\begin{align}
\ell (\bt) =&
  \nonumber
\sum_{i=1}^N \Big(\sum_{k=1}^{K-1}  \bigl(1-D_i\bigr) \bga_k^\T \bphi_e(\bx_i) I(y_i=k)
+    D_i \log\big[ 1 +  \sum_{k=1}^K \pi_k \{e^{\bga_k^\T \bphi_e(\bx_i)}-1 \} \big]\Big)\\
& \label{empirical_profile_log-likelihood}
 - \sum_{k=1}^{N}  \log \big[1+ \sum_{k=1}^K\lambda_k \{ e^{\bga_k^\T \bphi_e(\bx_i)} - 1 \}\big].
\end{align}

Given  $\ell(\bt)$, the maximum EL estimator (MELE)
 of $\bt$ is defined as
 \bas
\hat \bt
 = \arg \max_{\bt } \ell (\bt) .
 \eas
Substituting $\hat \bt$ into \eqref{p-mle} and \eqref{equ-lambda12} yields the MELE $\hat p_i$ of $p_i$. Accordingly, the MELEs for $F_0$ and $F_k$ are
 \begin{equation*}
 \hat F_0(\bx) = \sum_{i=1}^N \hat p_i I(\bX_i\leq \bx) \quad
 \mbox{and}\quad
 \hat F_k(\bx) = \sum_{i=1}^N \hat p_i e^{ \hat \bga_k^\T \bphi_e(\bX_i) } I(\bX_i\leq \bx), \;\; k=1, 2,\dots ,K.
\end{equation*}
 The explicit form of $\hat{\bt}$ is generally unknown. In the next subsection, we present an EM algorithm
 to numerically compute $\hat{\bt}$.

\subsection{EM Algorithm}
\label{Parameter_estimation}
Let $\mathcal{X}=\mathcal{L} \cup \mathcal{U}$ denote all the observed data, and
let $\{y_j^*: n+1\le j\le n+m\}$   be the latent labels for
the test  data.
If these labels were observed, the corresponding complete log-EL would be
\begin{align*}
% \hspace{-1in}
\ell^c(\bT) =&
 \sum_{k=1}^{N}  \log (p_i)
+  \sum_{i=1}^n \sum_{k=1}^{K-1}   \bga_k^\T \bphi_e(\bx_i) I(y_i=k)
+  \sum_{j=  n+1}^{N}     I(y_j^*=0) \log(1-\sum_{k=1}^K \pi_k)   \\ &
+  \sum_{j=  n+1}^{N} \sum_{k=1}^K    I(y_j^*=k) \log(\pi_k)  +  \sum_{j=  n+1}^{N} \sum_{k=1}^K I(y_j^*=k) \bga_k^\T \bphi_e(\bx_j),
\end{align*}
where $\bT=(\bga,\bpi,p_1,\dots , p_N)$.
Our EM algorithm is constructed  based on $\ell^c(\bT)$.

The core of the EM algorithm is the iterative EM procedure, which consists of an E-step and an M-step. Let $\bT^{(r)}$ denote the value of $\bT$ after the $r$-th EM iteration, with $r = 0, 1, 2, \dots$. When $r = 0$, $\bT^{(0)}$ represents an initial value of $\bT$.

\subsubsection*{E-step: Calculate
$
\mathcal{M} (\bT|\bT^{(r)})=\e\left\{\ell^c(\bT)|\mathcal{X}, \bT^{(r)}\right\}
$.} 

Given $\mathcal{X}$ and $ \bT^{(r)}$, for $j =n+1,\dots , N$ and $k=0, 1, 2, \dots , K$, the conditional expectation of  $I(y_j^*=k)$,
$\e\{ I(y_j^*=k) | \mathcal{X}, \bt^{(r)}  \} $,  is computed as
\begin{align}
\label{weighted_function_1k}
w_{jk}^{(r+1)}
=&
\frac{\pi_k^{(r)} e^{  \bga_k^{(r)\T} \bphi_e(\bx_j)}     }
{1+\sum_{k=1}^K \pi_k^{(r)}\{ e^{  \bga_k^{(r)\T} \bphi_e(\bx_j)} -1\}  }, \quad 1\le k \le K,\\
 w_{j0}^{(r+1)}  =& \label{weighted_function_0}   1- \sum_{k=1}^K w_{jk}^{(r+1)}.
\end{align}
Then, $\mathcal{M} (\bT|\bT^{(r)})$ becomes
\begin{align*}
\mathcal{M}(\bT|\bT^{(r)})
=&
\sum_{k=1}^{N}  \log (p_i)
+  \sum_{i=1}^n \sum_{k=1}^{K-1}    \bga_k^\T \bphi_e(\bx_i) I(y_i=k)  +   \sum_{j=  n+1}^{N}      w_{j0}^{(r+1)} \log(1-\sum_{k=1}^K \pi_k)   \\
&
+  \sum_{j=  n+1}^{N} \sum_{k=1}^K  w_{jk}^{(r+1)} \log(\pi_k)  +  \sum_{j=  n+1}^{N} \sum_{k=1}^K  w_{jk}^{(r+1)} \bga_k^\T \bphi_e(\bx_j).
\end{align*}

\subsubsection*{M-step: Update $\bT$ from $\bT^{(r)}$  to $\bT^{(r+1)}$ by 
}
$$
\bT^{(r+1)}=\arg\max_{\bT}
\mathcal{M}(\bT|\bT^{(r)})\quad\mbox{subject to the constraints in (\ref{constraints}).}
$$

Recall $n_k=\sum_{i=1}^nI(y_i=k)$, $k=0,1,\dots ,K-1$. Let $n_K=0$  and define
\begin{align*}
\mathcal{M}^{(r+1)}(\bga)=& \nonumber
\sum_{i=1}^n \sum_{k=1}^{K-1}    \big\{\alpha_k^*+ \bbeta_k^{\T}\phi(\bx_i)\big\} I(y_i=k)
 + \sum_{j=  n+1}^{N} \sum_{k=1}^K  w_{jk}^{(r+1)} \big\{\alpha_k^*+ \bbeta_k^{\T}\phi(\bx_j)\big\}\\
&-\sum_{i=1}^N\log\left\{ 1+\sum_{k=1}^K e^{\alpha_k^*+ \bbeta_k^{\T}\phi(\bx_i)} \right\},
\end{align*}
where
  \ba
  \label{alphastar}
\alpha_k^{\ast}=\alpha_k+\log\left(\frac{n_k+\sum_{j=n+1}^N w_{jk}^{(r+1)}}{n_0+\sum_{j=n+1}^N w_{j0}^{(r+1)}}\right), \quad 1\le k\le K.
  \ea
In Section 2.1 of the supplementary material, we show that $\bT^{(r+1)}$ is  computed as
  \begin{align}
\label{bga1}
\bga^{(r+1)}&=\arg \max_{\bga}\mathcal{M}^{(r+1)}(\bga),\\
\label{bga2}
  \pi_k^{(r+1)}&=\frac{1}{m}\sum_{j=n+1}^{n+m}w_{jk}^{(r+1)},\quad \text{for}\ k=1,2,\dots,K,\\
  \label{bga3}
p_i^{(r+1)}&=N^{-1}\left[1+\sum_{k=1}^K\exp\left\{\alpha_k^{\ast(r+1)}+\bbeta_k^{(r+1)\T}\bphi(\bx_i)\right\}\right]^{-1},
  \end{align}
 where  $\alpha_k^{\ast(r+1)}$ is given in \eqref{alphastar} with $\alpha_k$ replaced by $\alpha_k^{(r+1)}$.
It is worth mentioning that the objective function $\mathcal{M}^{(r+1)}(\bga)$ is proportional to the weighted log-likelihood of a multinomial logistic regression model with $K+1$ classes. Hence, $\bga^{(r+1)}$ can be readily obtained by fitting a multinomial logistic regression, which is supported by most software, for example, the \texttt{glmnet} function in the R package \texttt{glmnet}. Further details are provided in Section 2.1 of the supplementary material.

\begin{algorithm}[!ht]
  \caption{EM Algorithm for Parameter Estimation}\label{algorithm:spel-em}
  \begin{algorithmic}
    \Input  Labeled data \(\mathcal{L} = \{( \boldsymbol{x}_i, y_i)\}_{i=1}^n\); Unlabeled data $\mathcal{U} = \{\bx_i\}_{i=n+1}^{n+m}$.
    \Output  Estimates of $\bT$.
    \Initial Set $r=0$, $\bpi^{(0)}$, $\bga^{(0)}$
    \While{not converged}
      \State \textbf{E-step:}
     \quad  Compute $w_{jk}^{(r+1)}$'s using \eqref{weighted_function_1k}--\eqref{weighted_function_0};
      \State \textbf{M-step:}
   \quad Compute $\bT^{(r+1)}$ using \eqref{bga1}--\eqref{bga3}.
    \EndWhile
    \State Output the estimates.
  \end{algorithmic}
\end{algorithm}

Combining the E-step and M-step leads to the pseudocode for the EM algorithm, presented in Algorithm~\ref{algorithm:spel-em}.
The following proposition shows that
log-EL
$
\tilde\ell=\tilde\ell(\bT)
$ in \eqref{log_likelihood_func} does not  decrease after each iteration.
\begin{proposition}
  \label{theorem.em}
For the EM algorithm in \eqref{algorithm:spel-em},   we have
  $
  \tilde\ell(\bT^{(r+1)})
  \geq
  \tilde\ell(\bT^{(r)})
  $
   for $r\geq1$.
  \end{proposition}

%The proof of Proposition~\ref{theorem.em} is provided in Section 4 of the supplementary material.
We make two remarks about the EM algorithm. First, note that $\tilde \ell(\bT)$ under the constraints in \eqref{constraints} satisfies $\tilde \ell(\bT)\le 0$. With this result, Proposition~\ref{theorem.em} ensures that the EM algorithm converges to at least a local maximum for a given initial value $\bT^{(0)}$. To improve the chance of reaching the global maximum, we recommend using multiple initial values to explore the likelihood surface. Second, in practice, the algorithm may be terminated when the increase in the log-EL after an iteration is less than a prescribed tolerance, e.g., $10^{-5}$.

 \subsection{Asymptotic Properties}
\label{Statistical_Inference}
In this section, we investigate the limiting behavior of the proposed MELEs $\hat\bt$ and conduct inference on the mixture proportions $\pi_k$ in the test data. Based on the profile log-EL function in \eqref{empirical_profile_log-likelihood}, the empirical log-likelihood ratio (ELR) function for $\pi_k$, $k=0,1,\dots,K$, is defined as
$$
R_{N,k}(\pi_k)=2\left\{\ell(\hat\bt)
-\ell(\hat\bt_{k})
\right\},
$$
where $\hat\bt_{k}$ is the MELE of $\bt$ with $\pi_k$ held fixed.
The estimator $\hat\bt_{k}$ can be obtained by a slight modification of the M-step in Algorithm~\ref{algorithm:spel-em}. Details are provided in Section~2.2 of the supplementary material.

\begin{theorem}
\label{thm:thm1}
Under Assumptions~\ref{ass:ass1}--\ref{ass:ass2} and Conditions~C1--C3 in the Appendix, as $N \to \infty$:
\begin{enumerate}
  \item[\rm{(i)}] $\sqrt{N}(\hat{\bt}-\bt^{o}) \;\; \overset{d}{\longrightarrow} \; N\left(\bzero, \bSigma \right)$, where $\bSigma$ is defined in \eqref{asymptotic_variance} in the Appendix;
  \item[\rm{(ii)}] $R_{N,k}(\pi_k^o) \;\; \overset{d}{\longrightarrow} \; \chi_1^2$, for $k = 0,1,\dots ,K$;
  \item[\rm{(iii)}] The stochastic process $\sqrt{N}\{\hat F_k (\cdot)-F_k(\cdot) \}$ converges weakly to a mean-zero Gaussian process for each $k = 0,1,\dots,K$.
\end{enumerate}
\end{theorem}

Part~(ii) of Theorem~\ref{thm:thm1} provides the theoretical basis for constructing confidence intervals for the mixture proportion $\pi_k$, $k=0,1,\dots,K$. A $100(1-\alpha)\%$ EL ratio-based confidence interval for $\pi_k$ is given by
\begin{equation}
\label{elr.ci}
\{\pi_k : R_{N,k}(\pi_k) \le \chi_{1,1-\alpha}^2 \},
\end{equation}
where $\chi_{1,1-\alpha}^2$ denotes the $100(1-\alpha)\%$ quantile of the chi-square distribution with one degree of freedom. This method addresses an important gap in the existing literature, which often assumes $\bpi$ is known or provides only point estimates.

\section{Classification with an Approximately Optimal Classifier}
\label{Classification_under_SPEL_Framework}
The proposed MELE $\hat{\bt}$ plays an important role in our classification task. This section explains its application in constructing a classifier for the test data. As discussed in \cite{tian2024neyman}, the impact of misclassification can vary greatly across applications. For example, in loan outcome prediction, where possible results include default, full repayment, and late payment,
misclassifying a high-risk default as ``fully paid" can cause substantially greater financial loss than mistakenly flagging a reliable borrower as a default risk. This asymmetry in costs underscores the importance of learning methods that account for varying error severities.

Following \cite{tian2024neyman}, we consider a cost-sensitive classification problem for the test data. For a classifier $\mathcal{C}$ applied to the test set, the cost-sensitive loss is defined as
\begin{equation}
\label{cs_classification_problem}
\operatorname{Loss}(\mathcal{C}) = \sum_{k=0}^K \sum_{j \ne k} q(k, j) \cdot \pi_k \cdot P_{\text{te}}(\mathcal{C}(\bX) = j| Y = k).
\end{equation}
Here, $q(k, j)$ represents the user-specified cost of misclassifying a sample from true class $k$ as class $j$ ($j \ne k$), with $0 < q(k, j) < \infty$. When all $q(k, j)$ are equal for $k \ne j$, the problem reduces to the standard (uniform-cost) misclassification setting.

The optimal classifier that minimizes \eqref{cs_classification_problem} admits an explicit form, determined by the misclassification costs and the posterior probabilities $\{P_{\text{te}}(Y = k | \bX = \bx)\}_{k=0}^K$. The result is formally stated in the following lemma.
\begin{lemma}
  \label{general_bayes_optimal_classifier}
The classifier
\begin{equation}
\label{expr_optimal_classifier}
{{\mathcal{C}}}_{opt} (\bx) = \arg \min _{j\in \{0,1,\dots , K\}}\left\{\sum_{k \neq j} q(k,j) P_{\rm{te}}(Y=k | \bX=\bx)\right\}
\end{equation}
minimizes \eqref{cs_classification_problem} among all classifiers.
When the cost $q(k,j)$ is a constant for all $k\neq j$, the optimal classifier ${{\mathcal{C}}}_{opt}$ reduces to the commonly used Bayes classifier
\begin{equation}
\label{expr_bayes_classifier}
{{\mathcal{C}}}_{opt} (\bx)  = {\arg \max _{k\in \{0,1,\dots , K\}}}  P_{\rm{te}}(Y=k | \bX=\bx).
\end{equation}
\end{lemma}
As shown in Lemma \ref{general_bayes_optimal_classifier}, the posterior probabilities $\{P_{\text{te}}(Y = k \mid \bX = \bx)\}_{k=0}^K$ are fundamental to constructing the optimal classifier. Under model~\eqref{drm}, applying Bayes' rule with the conditional density $f_k(\bx) = P_{\text{te}}(\bx \mid Y = k)$ yields, we obtain for each $k = 0, 1, \dots, K$:
\begin{align}
\label{eq:bayes-posterior}
  \mathcal{C}_k(\bx;\bt):= P_{\text{te}}( Y = k|\bX=\bx)=\frac{\pi_k f_{k}(\bx)}{  \sum_{j=0}^K \pi_j f_{j}(\bx)}=\frac{\pi_k e^{\bga_k^{\T}\bphi_e(\bx)}}{\sum_{j=0}^{K}\pi_j e^{\bga_j^{\T}\bphi_e(\bx)}},
\end{align}
where in the last step, we have used $\pi_0 = 1 - \sum_{k=1}^K \pi_k$ and $\bga_0=\bzero$.
This expression for the posterior probability $P_{\text{te}}( Y = k|\bX=\bx)$ in \eqref{eq:bayes-posterior} highlights the value of the DRM beyond identifiability.

Given the MELE $\hat{\bt}$ and setting $\hat{\bga}_0 = \bzero$, a natural estimator of \eqref{eq:bayes-posterior} is
$ \mathcal{C}_k(\bx;\hat\bt)$.
The following theorem shows that the $L_1$-distance between $ \mathcal{C}_k(\bx;\hat\bt)$ and
$ \mathcal{C}_k(\bx;\bt^o)$ is of order $N^{-1/2}$.

\begin{theorem}
  \label{Bayes_theorem}
Assume the same conditions as in Theorem \ref{thm:thm1}.
We have
$$
\int \Big|\mathcal{C}_k(\bx;\hat\bt)-\mathcal{C}_k(\bx;\bt^o)\Big|P_{\rm{te}}(\bx)d\bx=O_p(N^{-\frac{1}{2}}),~~k=0,\ldots,K.
$$
\end{theorem}
%The proof of Theorem \ref{Bayes_theorem} is given in Section~7 of the supplementary material.
This theorem implies that $\mathcal{C}_k(\bx;\hat\bt)$ converges to $\mathcal{C}_k(\bx;\bt^o)$ as $N \to \infty$. Therefore,
substituting $\mathcal{C}_k(\bx;\hat\bt)$ into \eqref{expr_optimal_classifier} yields an approximately optimal classifier.

\section{Numerical Studies }
\label{Numerical_studies}
In this section, we use simulations to evaluate the performance of the proposed method in point estimation and confidence interval estimation of the $\pi_k$'s, as well as in classifying test observations. We then apply the method to a real-world dataset on phone prices to demonstrate its practical utility. Throughout both the simulation studies and the real-data analysis, we assume a constant cost $q(k,j)$ for $k\neq j$,
under which the optimal classifier is given in \eqref{expr_bayes_classifier}.
Using \eqref{eq:bayes-posterior},
the approximately optimal classifier is
\begin{equation}
\label{simulation_appro_optimal_classifier}
{\widehat{\mathcal{C}}}_{opt} (\bx) = \arg \max _{j\in \{0,1,\dots , K\}}\mathcal{C}_k(\bx;\hat\bt).
\end{equation}

\subsection{Simulation Study\label{Simulation_study}}

In our simulation study, we set $K=3$ and take $\bphi(\bx) = \bx$ in model~\eqref{drm} as the basis for the proposed method. Each distribution $F_k$ ($k = 0, 1, 2, 3$) follows a multivariate normal distribution $N(\boldsymbol{\mu}_k, \bI_{6})$, where the mean vectors are $\boldsymbol{\mu}_0 = (0, 0, 0, 0, 0, 0)^\T$, $\boldsymbol{\mu}_1 = (1, 1, 0, 2, 0, 0)^\T$, $\boldsymbol{\mu}_2 = (-1, -2, -1, 2, 0, 0)^\T$, and $\boldsymbol{\mu}_3 = (0, -1, -1, 1, 0, 0)^\T$, and $\bI_{6}$ is the $6 \times 6$ identity matrix. Under this specification, model~\eqref{drm} holds. We generate a training dataset of $n = 1200$ samples, consisting of $n_0$ observations from $F_0$ and $n_1 = n_2$ observations from $F_1$ and $F_2$, respectively. To investigate potential label shift between the training and test datasets, we consider two values for the ratio $n_0/n$: 1/2 and 1/3, which correspond to the presence and absence of label shift in the observed classes, respectively. The test dataset contains $m = 1200$ observations drawn from a mixture of $F_0, \ldots, F_3$ with mixture proportions $\pi_0 = 0.2$ and $\bpi = (\pi_1, \pi_2, \pi_3)^\T = (0.2, 0.2, 0.4)^\T$. Each simulation scenario is repeated 1000 times.

\noindent{\bf Mixture Proportion Estimation}
In this part, we evaluate the performance of the proposed point estimator and confidence intervals for the $\pi_k$'s. Our assessment focuses on two main tasks: 1) Examining the root mean square error (RMSE) and relative bias (RB) of the proposed MELE for $\bpi$, in comparison with the PULSE method\footnote{Implemented in {\tt Python}; available at \href{https://github.com/acmi-lab/Open-Set-Label-Shift}{https://github.com/acmi-lab/Open-Set-Label-Shift}} introduced by \citet{garg2022domain}. The PULSE method represents a recent advancement in the literature, offering improved performance over earlier approaches such as \citet{blanchard2010semi} and related derivatives, which suffer from computational intractability and error accumulation, as discussed in Section~\ref{challenges_related_work}. 2) Evaluating the coverage probability (CP) of the proposed confidence intervals
in \eqref{elr.ci} for the $\pi_k$'s. In our simulations,
we use a nominal level of $95\%$.

Simulation results  are summarized in Table \ref{tab-pi_rmse_bias_1200_dim6_3+1}.
We observe that the MELE performs very well: RBs are negligible ($\le$ 1.0\%) across all components, and CPs remain close to the nominal 95\% level under all scenarios. In contrast, the PULSE estimator shows non-negligible RB (around 12.5\% for $\pi_1$ and 6.0\% for  $\pi_3$) and higher RMSE across all settings. These results suggest that the proposed method provides consistent estimation of $\bpi$, while PULSE not only shows systematic bias but also cannot construct confidence intervals for the $\pi_k$'s.

\begin{table}[h]
  \caption{
Simulated relative bias (RB, $\times$ 100), root mean square error (RMSE, $\times$ 100), and coverage probability ($\rm{CP}$, $\times$ 100)
of the MELE and PULSE estimators for $\bpi$.
     \label{tab-pi_rmse_bias_1200_dim6_3+1} }
\vspace{2ex}
    \begin{adjustbox}{center}
        \centering
    \begin{tabular}{cccccccccc}
    \hline
   $n_0/n$ & ${\bpi}$ & & \multicolumn{2}{c}{PULSE }& & \multicolumn{3}{c}{MELE }\\
  % \cline{4-5}
   \cmidrule(r){4-5} \cmidrule(r){7-9}
   & & &  RB & RMSE  & & RB &  RMSE & $\rm{CP}$ \\
    \hline
    1/3
    & ${\pi}_1$  && 12.5 & 4.4 && 0.0 & 1.8 & 95.1\\
   & ${\pi}_2$ && -5.0 & 5.1 && -1.0 & 3.1 & 93.3\\
   & ${\pi}_3$ && 6.0 & 10.1 &&   0.75 & 4.0 & 94.4 \\
   \cline{2-9}
    1/2
     & ${\pi}_1$ && 12.5 & 4.7&& 0.0 & 1.9 & 94.0\\
   & ${\pi}_2$ && -4.0& 5.5 &&   -1.0 & 3.3 & 93.8 \\
    & ${\pi}_3$ && 5.0& 10.5 &&   0.75 & 4.0 & 95.1 \\

   \hline
\end{tabular}
\end{adjustbox}
\end{table}

\noindent{\bf Classification Accuracy}
As a practical application of our proposed framework, we consider classification using the approximately optimal classifier in \eqref{simulation_appro_optimal_classifier}. We evaluate the performance of our method under the experimental settings described at the beginning of this section. Additionally, we compare our approach with the multinomial PU method (Mul-PU) proposed by \citet{zheng2023high}. Unlike our method, Mul-PU relies on prespecified proportions
 $\bpi$ rather than estimating them from the observed data, placing it in the first class of methods reviewed in Section~\ref{challenges_related_work}.

\begin{figure}[!ht]
    \centering
    \includegraphics[width=0.9\textwidth]{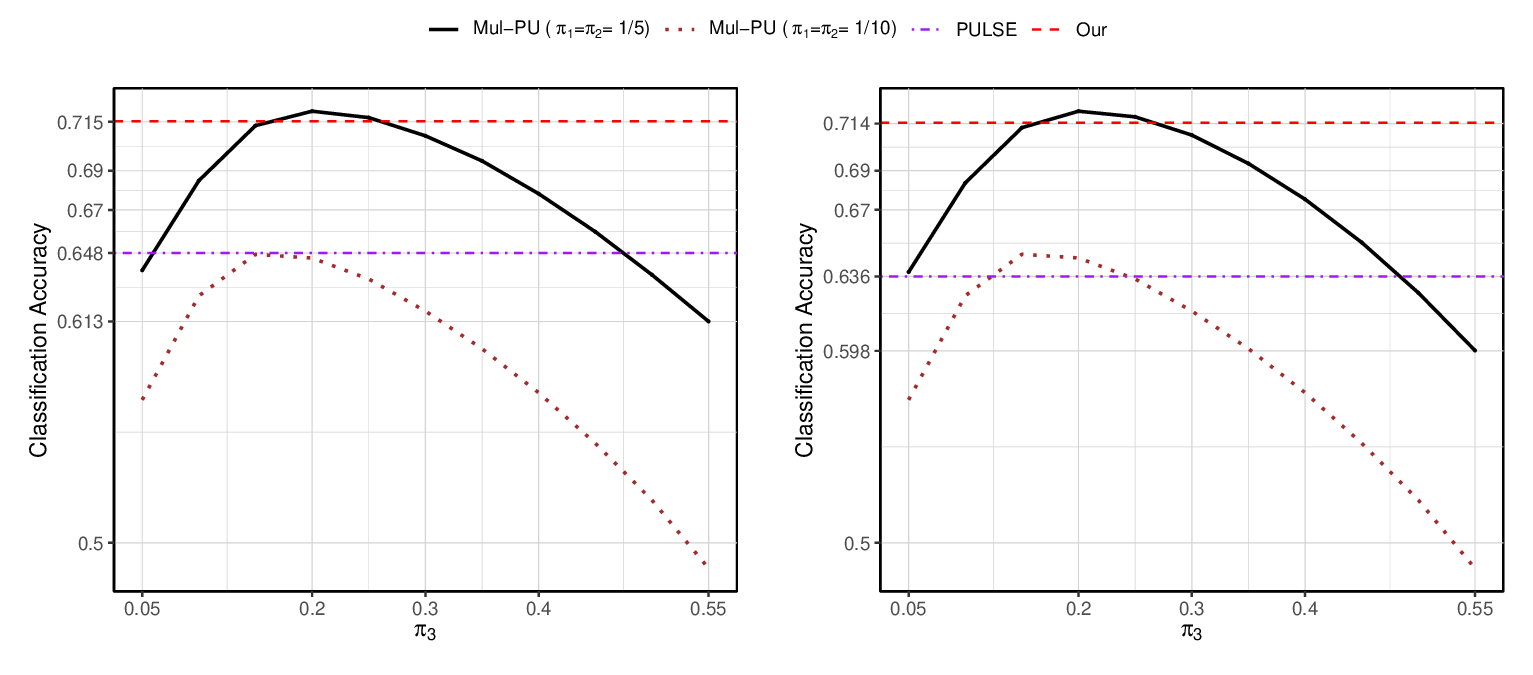}
    \caption{
      Simulated classification accuracies under varying values of $\pi_3$. Results are shown for PULSE (purple, dot-dashed), our method (red, dashed), and Mul-PU under two specifications: $\pi_1 = \pi_2 = 1/5$ (black, solid) and $\pi_1 = \pi_2 = 1/10$ (brown, dotted). The left and right panels correspond to $n_0/n = 1/3$ and $n_0/n = 1/2$, respectively.}
    \label{Fig-3+1_compare_lili_simulated}
    \end{figure}

To avoid overfitting, we generate a separate validation dataset of size $m^* = 1200$ from the test distribution. All classifiers are evaluated on this validation set to assess classification accuracy. We examine two configurations of Mul-PU: one with correctly specified values \(\pi_1 = \pi_2 = 1/5\), and another with misspecified values \(\pi_1 = \pi_2 = 1/10\). We further investigate the influence of \(\pi_3\) varying within \([0.05, 0.55]\) on classification accuracy. Figure \ref{Fig-3+1_compare_lili_simulated} displays the empirical classification accuracies of Mul-PU, PULSE, and our method. The accuracy of Mul-PU shows a clear increasing trend followed by a decline in both scenarios, with markedly better performance under correct specification of \(\pi_1\) and \(\pi_2\). In comparison, our method achieves an accuracy of 0.715 when \(n_0/n = 1/3\), outperforming PULSE by approximately 7\%. All methods exhibit similar trends when \(n_0/n = 1/2\), indicating their feasibility under label shift.

In summary, when model \eqref{drm} holds, our method, owing to its consistent estimation of model parameters, demonstrates superior and more robust classification performance.

  \subsection{Real Data Analysis}
  \label{Real_Data_Analysis}
  This section demonstrates the proposed methodology using a real data application.
  We consider the Mobile Phone Price dataset from Kaggle\footnote{Available at \href{https://www.kaggle.com/datasets/iabhishekofficial/mobile-price-classification}{https://www.kaggle.com/datasets/iabhishekofficial/mobile-price-classification}}, which contains 20 features and an ordinal label indicating the phone's price range from low to very high cost (values in {0, 1, 2, 3}). Each class contains 500 observations.
  The features include properties such as the memory size and the phone's weight; see Table S1 in the supplementary materials for the full list of the features.

We begin by pre-processing the dataset, centering and standardizing each covariate. Class 3 (high-end phones) is treated as the novel class in the test data. The training data is constructed using 50\% of the data from each of classes 0, 1, and 2, yielding $n = 750$ samples. The prediction set consists of the remaining 50\% from classes 0-2 and all observations from class 3, resulting in $m = 1250$ samples, with proportions $\pi_0 = \pi_1 = \pi_2 = 0.2$ and $\pi_3 = 0.4$.

\begin{figure}[!ht]
      \centering\includegraphics[width=0.8\textwidth]{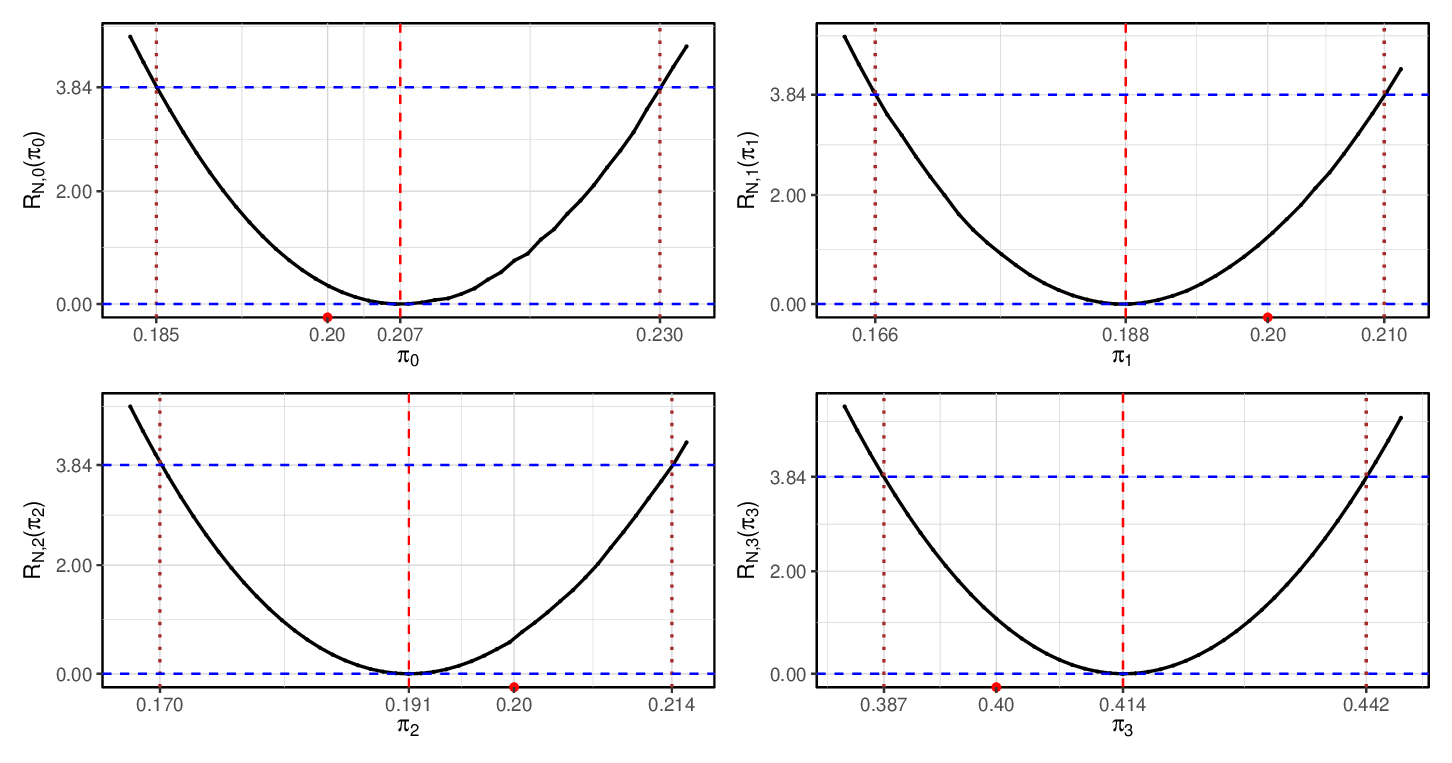}
      \caption{
      Plots of the EL ratio functions $R_{N,k}(\pi_k)$ versus $\pi_k$ for $k=0,1,2,3$.
      % Red points denote true values, vertical dashed lines indicate MELE. Blue horizontal lines show the 95\% Chi-square quantile; brown dotted lines mark 95\% confidence intervals.
        }
      \label{Fig-R_N_binary_real_data_full}
      \end{figure}

We then examine the estimation and inference results for the mixture proportion $\bpi$ using the EL ratio functions $ R_{N,k}(\pi_k)$ , for $ k = 0, 1, 2, 3,$ as illustrated in Figure~\ref{Fig-R_N_binary_real_data_full}.
Here, we use the full prediction set as the test data.
The MELEs are $\widehat{\pi}_0 = 0.207$, $\widehat{\pi}_1 = 0.188$, $\widehat{\pi}_2 = 0.191$, and $\widehat{\pi}_3 = 0.414$, each lying close to their respective true values $(0.2, 0.2, 0.2, 0.4)^{\T}$. The 95\% confidence intervals, namely [0.185, 0.230], [0.166, 0.210], [0.170, 0.214], and [0.387, 0.442], all contain the corresponding true value of $\pi_k$. In the figure, vertical dashed red lines mark the MELE, the blue horizontal line (at 3.84) represents the 95\% quantile of the $\chi^2_1$ distribution, and brown dotted lines indicate the confidence bounds. Notably, these intervals do not cover the first three proportion estimates from the PULSE method, as reported in Table~\ref{tab-pi_point_cr_est}.

      \begin{table}[h]
        \caption{
          Point estimates (PE) and 95\% confidence intervals (CI) for mixture proportions under MELE, with comparative point estimates from PULSE.
      % All results are rounded to  third decimal place.
           \label{tab-pi_point_cr_est} }
      \vspace{2ex}
          \begin{adjustbox}{center}
              \centering
          \begin{tabular}{cccccc}
          \hline
          Mixture & True & \multicolumn{2}{c}{MELE}& & {PULSE }\\
        % \cline{4-5}
         \cmidrule(r){3-4}
         Proportion & Value &  PE & CI  & & PE   \\
          \hline
           ${\pi}_0$  & 0.2 & 0.207 & [0.185,0.230] &  & 0.112 \\
           ${\pi}_1$  & 0.2 & 0.188 & [0.166,0.210] && 0.133 \\
         ${\pi}_2$ & 0.2 & 0.191 & [0.170,0.214] && 0.368 \\
          ${\pi}_3$ &0.4 & 0.414 & [0.387,0.442] &&   0.388 \\
         \hline
      \end{tabular}
      \end{adjustbox}
      \end{table}

\begin{figure}[!ht]
        \centering
        \includegraphics[width=0.66\textwidth]{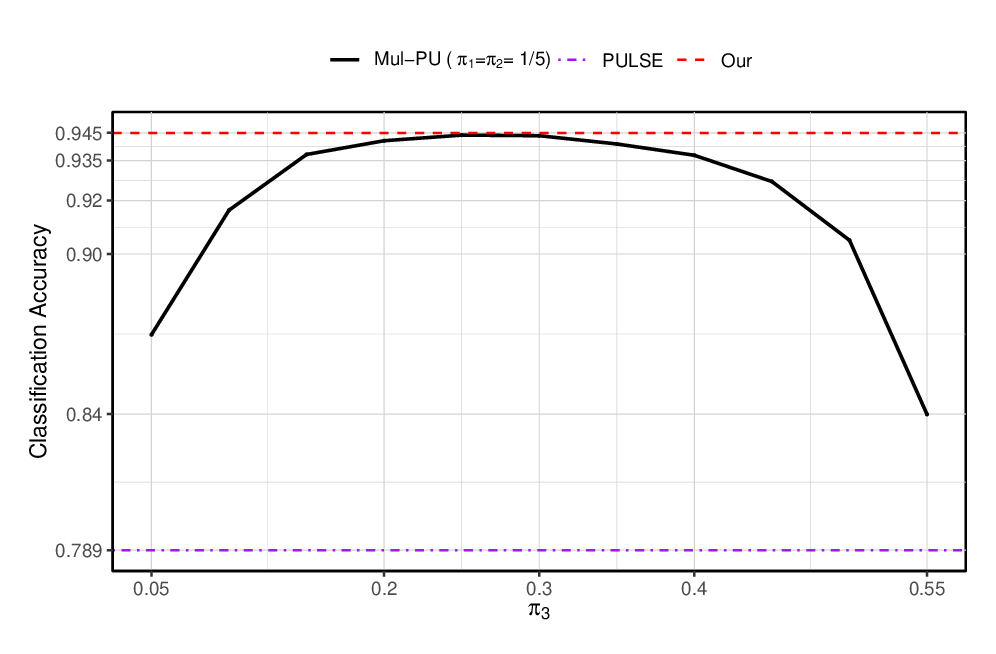}
        \caption{
          Average classification accuracies comparing our method (red, dashed), PULSE (purple, dot-dashed), and Mul-PU (black, solid; $\pi_1=\pi_2=1/5$) under different $\pi_3$ on the Mobile Phone Price dataset. 
       }
        \label{Fig-real_data_4_lili}
        \end{figure}

      Finally, we evaluate the classification performance of PULSE, Mul-PU, and our method. To do so, the prediction set is further randomly split in a 70/30 ratio into test and validation subsets. The model is trained on the combined training and test sets, and its classification performance is assessed on the validation set¡ªa process repeated across 100 random partitions. Figure \ref{Fig-real_data_4_lili} plots the average empirical accuracies of three methods across these 100 repetitions. With an accuracy of 0.945, our method surpasses all trial values of $\pi_3$ when applied to Mul-PU.
       PULSE achieves an accuracy of 0.789, which is the lowest among all methods compared.

\section{Discussion}
\label{Discussion}
This paper focuses on OSLS problem, where a novel class may appear in the test data. To address the identifiability challenge, we employ a DRM and propose a MELE for estimating the class proportions in the test data, along with EL ratio based confidence intervals. An EM algorithm is developed for numerical implementation. Theoretically, we establish the asymptotic properties of the proposed inference procedures, which provide a foundation for both point estimation and confidence interval construction. Furthermore, we assign labels to the test data by constructing an approximation to the optimal classifier based on the estimated posterior probabilities, and we show that it achieves a convergence rate of $N^{-1/2}$.

Our work opens several directions for future research. For instance, exploring how penalized empirical likelihood methods can be effectively applied in high-dimensional feature spaces is a worthwhile avenue. In addition, considering the potential misspecification of the DRM, and noting that the conditional distribution of $\bX$ given $Y$ among observed classes can be learned through nonparametric approaches, one could relax the DRM assumption between observed classes and instead employ machine learning techniques (e.g., neural networks) to estimate the density ratio. We leave these extensions for future investigation.

% \section{Acknowledgments}
%The authors thank the anonymous reviewers for their valuable suggestions.  

\setcounter{section}{1}

\section*{\label{Appendix2}Appendix: Form of $\bSigma$ and Regularity Conditions}

Recall the notation:
$\bga_k^{o\T}=(\alpha_k^{o},\bbeta_k^{o\T}),$ $c_k=\sum_{i=1}^nI(y_i=k)/N$ for $k=1,2, \dots, K-1$, and define
$\lambda_{k}^{o} = c_k + c\pi_k^{o}$, for $k=1,2, \dots, K-1$, and $\lambda_{K}^{o} = c\pi_K^{o}$.
We also introduce
 \begin{align*}
 A^{o}( \bx)
 =&
 1+ \sum_{k=1}^K\lambda_k^{o}\{ e^{\bga_k^{o\T}\bphi_e( \bx)}-1\},\quad
 B^{o}(\bx)
 =
 1+ \sum_{k=1}^K\pi_k^{o}\{ e^{\bga_k^{o\T}\bphi_e( \bx)}-1\}, \\
\bpi^{o}
 =&
 (\pi_1^o, \pi_2^o, \dots , \pi_K^o)^{\T}, \quad
 \blambda^{o}
 =
 (\lambda_1^o, \lambda_2^o, \dots , \lambda_K^o)^{\T},\\
 \bQ^{o}( \bx)
 =&
 \Big(  e^{\bga_1^{o\T}\bphi_e( \bx)}-1,  e^{\bga_2^{o\T}\bphi_e( \bx)}-1 ,\dots , e^{\bga_K^{o\T}\bphi_e( \bx)}-1 \Big)^{\T},\\
 \bS^{o}( \bx)
 =&
 \Big(  e^{\bga_1^{o\T}\bphi_e( \bx)},  e^{\bga_2^{o\T}\bphi_e( \bx)} ,\dots , e^{\bga_K^{o\T}\bphi_e( \bx)} \Big)^{\T}.
 \end{align*}

 The asymptotic variance matrix $ \bSigma$ is given by
 \ba
 \label{asymptotic_variance}
 \bSigma = \bW_*^{-1},
 \ea
 where
 \begin{align}
 \bW_* =& \label{w_*}-\left(\begin{array}{lll}
  \bW_{11}- \bW_{13} \bW_{33}^{-1} \bW_{31} &  \bW_{12}\\
  \bW_{21}  &  \bW_{22}
     \end{array}\right),
\end{align}
and the components matrices are specified as follows:
\begin{align*}
 \bW_{11} =&
  \e_0  \Big[ \frac{ \{{\blambda}^{o} \odot  \bS^o( \bX) \}^{\otimes 2} \otimes \{ \bphi_e( \bX)\}^{\otimes 2}}{
A^{o}( \bX )   } \Big]
-
c\e_0  \Big[ \frac{ \{{\bpi}^{o}\odot  \bS^o( \bX) \}^{\otimes 2} \otimes \{ \bphi_e( \bX)\}^{\otimes 2}}{
 B^{o}( \bX )   } \Big]\\
 & -  \e_0 \Big[  \diag \big\{ (\blambda^{o}-c \boldsymbol{\pi}^{o} )\odot \bS^o( \bX)\big\} \otimes  \{ \bphi_e( \bX)\}^{\otimes 2} \Big],
\\
\bW_{12} =&  \bW_{21}^{\T}=
c \e_0 \Big[  \diag\big\{  \bS^o( \bX)\big\} \otimes  \bphi_e( \bX)\Big]
\\
&\quad\quad\quad -c \e_0 \Big[\frac{   \big\{ {\bpi}^{o}\odot  \bS^o( \bX)\big\} \otimes  \{ \bphi_e( \bX)\bQ^{o\T}( \bX)\}}{B^o( \bX)} \Big],\\
\bW_{13}
=& \bW_{31}^{\T}=
\e_0 \Big[\frac{   \big\{ {\blambda}^{o}\odot  \bS^o( \bX)\big\} \otimes  \{ \bphi_e( \bX) \bQ^{o\T}(\bX)\}}{A^o( \bX)} \Big]
- \e_0 \Big[  \diag\big\{  \bS^o( \bX)\big\} \otimes  \bphi_e(\bX)\Big]
,\\
\bW_{22} =&
-   c
\e_0 \frac{   \{  \bQ^o( \bX  )\}^{\otimes 2} }{  B^{o }( \bX )}, \quad
\bW_{23}
=  \bW_{32}^{\T}= \bzero, \quad
\bW_{33} =
\e_0
\frac{ \{  \bQ^o( \bX  )\}^{\otimes 2}}{
A^{o }( \bX )}.
\end{align*}
Here, $\odot$ denotes the Hadamard (elementwise) product, $\otimes$ the Kronecker product, and for a vector $\bba$, $\bba^{\otimes 2} = \bba \bba^{\T}$. In addition, $\diag\{\bba\}$ denotes the diagonal matrix with the entries of $\bba$ on its diagonal.

The asymptotic results in Theorem \ref{thm:thm1} rely on the following regularity conditions:

\begin{itemize}
  \sloppy
      \item[C1.]
The function  $\e_0  [ \exp\{   \bbeta_k^\T \bphi(\bX) \}   ]  $ is finite
for  $\bbeta_k $ in a neighborhood of $\bbeta_k^o$ and $k=1, 2, \dots , K$;
\item[C2.] The matrix $\bW_*$ defined in  \eqref{w_*} is nonsingular;
\item[C3.]
 $\bt^o$ is an interior point of the parameter space of $\bt$.
\end{itemize}
Condition C1 ensures that, for $\bt$ in a neighborhood of the true value $\bt^o$, $\ell(\bt)$ can be well approximated by a quadratic form in $\bt - \bt^o$ with a negligible remainder. Conditions C2 and C3 are standard assumptions commonly used in establishing the asymptotic normality of MELEs in the literature.

%\section{Author contributions statement}
%Must include all authors, identified by initials, for example: S.R. and D.A. conceived the experiment(s),  S.R. conducted the experiment(s), S.R. and D.A. analysed the results.  S.R. and D.A. wrote and reviewed the manuscript.

\bibliographystyle{natbib}
\bibliography{references}

\end{document}